\documentclass[a4paper,12pt]{article}

\usepackage[a4paper]{geometry}
\usepackage[utf8]{inputenc} 
\usepackage{hyperref}
\usepackage{float}
\usepackage{amssymb}
\usepackage{amsmath}
\usepackage{amsthm}
\usepackage{url}
\usepackage{amsfonts}
\usepackage{graphicx, float, tabularx}
\usepackage{color}
\usepackage{enumerate}
\usepackage[american]{babel}
\usepackage{stackrel}
\usepackage{wrapfig}
\usepackage{subfigure}

\newcommand{\be}{\begin{equation}}
\newcommand{\ee}{\end{equation}}
\newcommand{\bea}{\begin{eqnarray}}
\newcommand{\eea}{\end{eqnarray}}

\renewcommand{\frac}[2]{{{\displaystyle #1}\over{\displaystyle #2}}}

\author{\textbf{Athanasios G. Tzikas} \thanks{E-mail: \href{mailto:tzikas@fias.uni-frankfurt.de}{\texttt{tzikas@fias.uni-frankfurt.de}} 
} 
\\[1ex]
\small Frankfurt Institute for Advanced Studies (FIAS)\\[-0.5ex]
\small Ruth-Moufang-Str.~1, D-60438 Frankfurt am Main, Germany\\[0.5ex]
\small Institut f\"{u}r Theoretische Physik, Goethe-Universit\"{a}t Frankfurt am Main\\[-0.5ex]
\small Max-von-Laue-Str.~1, D-60438 Frankfurt am Main, Germany\\[1ex]
}
\date{\today}
\title{\textbf{Bardeen black hole chemistry}}

\begin{document}
\maketitle

\begin{abstract}
\noindent  In the present paper we try to connect the Bardeen black hole  with the concept of the recently proposed black hole chemistry. We study  thermodynamic properties of the regular black hole with an anti-deSitter background. The negative cosmological constant $\Lambda$ plays the role of the positive thermodynamic pressure of the system. After studying the thermodynamic variables, we derive the corresponding equation of state and we show that a neutral Bardeen-anti-deSitter black hole has similar phenomenology to the chemical Van der Waals fluid. This is equivalent to saying that the system exhibits criticality and a first order   small/large  black hole phase transition reminiscent of the liquid/gas coexistence.


\end{abstract}

\thispagestyle{empty}
\newpage
\vspace{0.1cm}




\section{Introduction} 
Black holes are among the most fascinating and mysterious astrophysical objects, predicted theoretically  in the context of general relativity (GR). At the same time, GR brings about its own downfall by predicting singularities at the center of a black hole, due to the ignorance of quantum mechanical effects. A complete theory of quantum gravity is supposed to cure these problems but such a formulation is still an ongoing research.  In the 1970's, Hawking  showed that  black holes emit semiclassically thermal radiation like almost perfect black bodies \cite{Haw74}. For the standard Schwarzschild  black hole, the Hawking temperature is inversely proportional to its mass, signalling a negative heat capacity for the black hole along with its instability. In 1980's,  Hawking and Page \cite{HaP83} studied thermodynamical aspects of a black hole inside an anti-deSitter (AdS) background revealing that it can undergo a phase transition between pure radiation and a stable black hole, known also as Hawking-Page phase transition. After that, black holes in AdS background became of much theoretical interest until today. One of the most successful theories is the AdS/CFT duality  \cite{Mal99,Wit98}  which relates  AdS gravitational theories to  conformal field theories formulated on a boundary. Another recent interesting proposal  \cite{KuM14}  is the treatment of the negative   cosmological constant $\Lambda$ of an AdS background as a thermodynamic variable related to the positive pressure $P$ of the system through the relation 
\begin{equation} \label{press}
P=-\frac{\Lambda}{8\pi} \,.
\end{equation}
The introduction of a pressure term revealed new interesting properties for semiclassical black holes analogous to known chemical phenomena, such as $P-V$ criticality for charged AdS black holes like in Van der Waals gas \cite{KuM12}, triple points like  water \cite{AKM+14} or reentrant phase transitions like in nicotine/water mixture \cite{AKM13,FKM++14}. Furthermore, the mass of the black hole is identified with the chemical enthalpy $H$ inside the extended phase space \cite{KRT09}, rather than the internal energy $U$, through the relation
\begin{equation}
M\equiv H=U+PV \,,
\end{equation}
where $V$ is the conjugate variable of $P$ identified as the thermodynamic volume of the black hole \cite{Dol11,Dol+11}. Recently, it has been shown that the enthalpy approach is directly connected to the quasilocal energy path integral Euclidean approach \cite{LeZ18}. The interpretation of the mass as the total energy of the system, i.e., the energy U needed to form a black hole plus the energy PV needed to place it inside a cosmological background, makes the 1st law of black hole mechanics \cite{BCH73} consistent with the known Smarr relation \cite{Sma73} which connects extensive with intensive thermodynamic variables:
\begin{eqnarray} \label{1st_law}
\mathrm{d} M &=& T \mathrm{d} S + P \mathrm{d} V  \,,  \\ 
M &=& 2 T S - 2 P V   \,.  \label{Smarr}
\end{eqnarray}
This whole new scientific area is called black hole chemistry \cite{KuM17}, with the word 'chemistry' interpreted here as the black hole thermodynamics with $\Lambda\,$.

As mentioned above, the existence of singularities is  an interesting problem for  GR  but they can be avoided by effective approaches \cite{Dym02,NSS06,Nic09,MMN11,IMN13,NiS14,FKN16,SpS17D,Nic18,CMN15,WiN11}. Such effective theories cure the short-distance pathologies by introducing a minimal cut-off length, close to the Planck length, under which gravity cannot be tested, providing us with singularity-free black holes. 

One of the first models  of  regular  black hole is proposed by  Bardeen \cite{Bar86} in 1968 (see \cite{Ans08} for a review).  Bardeen black hole is a spherically symmetric solution  resulting from  Einstein equations in the presence of a nonlinear electromagnetic field coupled to matter \cite{AyG00}. Thus it is parametrized by two quantities; a mass $M$ and a  charge $q$ both having dimensions of length in geometric units ($c=G=1$). The Bardeen line element reads
\begin{equation} \label{metric}
\mathrm{d} s^2 = -\left( 1-\frac{2m(r)}{r} \right)  \mathrm{d} t^2 + \left( 1-\frac{2m(r)}{r} \right) ^{-1}  \mathrm{d} r^2 + r^2 \mathrm{d} \Omega^2 \,
\end{equation}
with
\begin{equation} \label{mass_param}
m(r) = M \left[ 1+ \left( \frac{q}{r}\right)^2  \right] ^{-3/2} 
\end{equation}
and $\mathrm{d} \Omega^2 =\mathrm{d} \theta ^2 + \sin ^2 \theta \mathrm{d} \phi ^2\,.$  It is our goal  to connect Bardeen black holes with  the concept of black hole chemistry. The procedure adopted here is strongly inspired by \cite{KuM12,GMK13}. The paper is organized as follows: in Sec.~\ref{BAdS_section} we  extend the work of \cite{AyG00} by including a negative cosmological constant in the gravitational action and we derive the Bardeen-anti-deSitter (BAdS) line element. In Sec.~\ref{BAdS_chemistry} we investigate the chemical properties of BAdS  black hole, i.e., the thermodynamics with a $\Lambda$-term,   and we connect it with the known chemical  Van der Waals fluid. Specifically, we derive the corresponding equation of state and we show that the black hole appears a first order phase transition between small/large black hole, along with a $P-V$ criticality with critical exponents identical to those of a Van der Waals gas. In Sec.~\ref{concl} we summarize our conclusions. We are working in Planck units where $\hbar=c=G=k_{\rm{B}}=1\,.$

\section{Bardeen-anti-deSitter black hole}
\label{BAdS_section}

The corresponding action with a negative $\Lambda$-term, resulting from the extension of \cite{AyG00}, reads
\begin{equation}
\mathcal{S}= \frac{1}{16\pi} \int \mathrm{d} ^4x \sqrt{-g} \left( R+ \frac{6}{l^2} - 4 \mathcal{L}(F) \right) \,,
\end{equation}
where $R$ is the Ricci scalar, $g$ is the determinant of the metric tensor, $l$ is the positive AdS radius connected with $\Lambda$ through the relation $\Lambda=-3/l^2$ and $\mathcal{L}(F) $ is a function of $F=\frac{1}{4} F_{\mu\nu}F^{\mu\nu}$ given by
\begin{equation} \label{L_term}
\mathcal{L}(F) = \frac{3 M}{ |q|^3} \left( \frac{\sqrt{4 q^2 F}}{1+\sqrt{4 q^2 F}} \right) ^{5/2} ,
\end{equation}
with $q$ being a $\mathrm{U}(1)$ charge and $F_{\mu\nu}$ the  field tensor.  Varying the above action with respect to the metric and the electromagnetic field, we get the following  field equations of motion:
\begin{eqnarray} \label{field_eq1}
G_{\mu\nu} - \frac{3}{l^2}  g_{\mu\nu} &=& 2 \left( \frac{\partial \mathcal{L}(F)}{\partial F} F_{\mu \lambda} F _{\nu}{}^{\lambda} - g_{\mu\nu} \mathcal{L}(F) \right) \,, \\ 
0 &=& \nabla_{\mu} \left( \frac{\partial \mathcal{L}(F)}{\partial F} F^{\mu\nu}  \right) \,, \label{field_eq2}
\end{eqnarray}
where $G_{\mu\nu}$ is the known Einstein tensor. We imply a static and spherically symmetric line element of the form \eqref{metric}. With this metric ansatz and with the help of \eqref{field_eq2}, we find that
\begin{equation}
F_{\mu\nu} = 2 \delta ^{\theta}{}_{[\mu} \delta ^{\phi}{}_{\nu]} \ q(r) \sin\theta \,.
\end{equation}
Using also the condition $\mathbf{dF} = \frac{\mathrm{d} q}{\mathrm{d} r}  \sin \theta \ \mathrm{d} r \wedge \mathrm{d} \theta \wedge \mathrm{d} \phi= 0\,,$ we conclude to  the relation  $q(r)=\mathrm{const.}=q$ and, hence, the  field strength is $F_{\theta\phi}=q \sin \theta$ with $F = \frac{q^2}{4r^4}\,$. Substituting these expressions in \eqref{L_term}, we get
\begin{equation}
\mathcal{L}(F)=\mathcal{L}(r) = \frac{3M q^2}{(r^2 + q^2)^{5/2}} \,.
\end{equation}
Now the above field equations of motion yield a solution for $m(r)$ that reads
\begin{equation}
m(r)= \frac{M r^3}{(q^2+r^2)^{3/2}} - \frac{ r^3}{2 l^2} \,,
\end{equation}
fixing the BAdS metric potential to have the form of
\begin{equation} \label{BAdS_potential}
f(r) = 1 - \frac{2M r^2}{(q^2 + r^2)^{3/2}} + \frac{r^2}{l^2} \,.
\end{equation}
The charge $q$ has dimensions of length and thus it can be interpreted alternatively as a minimal cut-off length that makes gravity ultraviolet self-complete \cite{SpS14}. From now on we will consider $q$ as the minimal length for the BAdS black hole rather than a charge. 

\begin{figure}[h!] 
\begin{center}
\includegraphics[width=0.58 \textwidth]{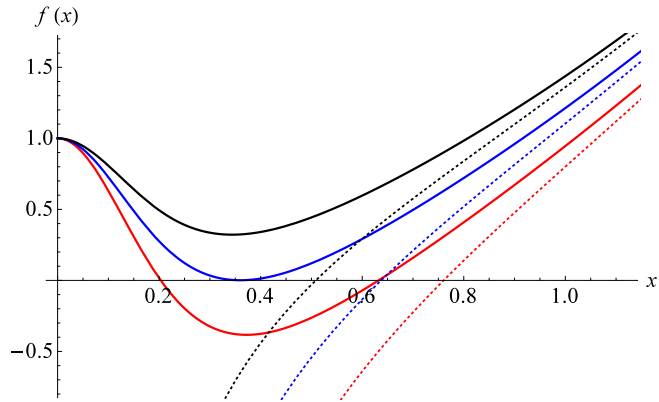}
\caption{The function $f(x)$ \textit{vs} $x$ for $Q=0.3\,.$ The solid curves with colors red, blue and black, correspond to $m=0.6\,$, $m=m_0 \approx 0.448 $ and $m=0.32$ respectively. The dotted lines correspond to $Q=0$ for each case and represent the usual Schwarzschild-anti-deSitter (SAdS) black hole.}
\label{fig:Bpotential}
\end{center}
\end{figure}

It would be more convenient to study the horizon structure in $l$ units and this can be achieved by writing  \eqref{BAdS_potential} as
\begin{equation} \label{BAdS2_potential}
f(x) = 1 - \frac{2 m x^2}{(Q^2+x^2)^{3/2}} + x^2 \,,
\end{equation}
where
\begin{equation} \label{dim_var}
x=r/l\,, \ \ \ \ m=M/l \,, \ \ \ \  Q=q/l \,.
\end{equation} 
 In Fig.~\ref{fig:Bpotential} we give the plot of \eqref{BAdS2_potential} for a fixed $Q$ and a varying $m\,$.
One sees the regularity at the center ($x=0$) of the black hole.
The parameter $Q$ plays the role of the "fictitious charge" analogous to the electric charge of the conventional Reissner-Nördstrom-AdS black hole, which is responsible for different causal structures \cite{CEJ+99,CEJM99}. Similar behaviour appears also the minimal cut-off length $\sqrt{\theta}$ in the noncommutative case \cite{NiT11}. 
Specifically, we get:
\begin{itemize}
\item Two horizons for $m>m_0$ (red solid curve in Fig.~\ref{fig:Bpotential}); one inner (Cauchy) $x_-$ and one outer (event) horizon $x_+\,.$ 
\item One degenerate horizon $x_0=x_-=x_+$ for $m=m_0\,$, representing the BAdS black hole with the smallest possible radius and mass (blue solid curve in Fig.~\ref{fig:Bpotential}).
\item No horizons for masses below $m_0$ (black solid curve in Fig.~\ref{fig:Bpotential}).
\end{itemize}
The values $x_0$ and $m_0$ of the degenerate black hole case can be determined from the solution of the system
\begin{equation}
f(x)=0=\frac{\partial f(x)}{\partial x} \,,
\end{equation}
which gives
\begin{eqnarray}
x_0 = \frac{\sqrt{-1+ \sqrt{1+24 Q^2}}}{\sqrt{6}} \,, \ \ \ \ m_0=\frac{\left( Q^2 + \frac{1}{6} \left( -1+\sqrt{1+24 Q^2} \right)  \right) ^{5/2}}{2Q^2 - \frac{1}{6} \left( -1+\sqrt{1+24 Q^2} \right) } \,.
\end{eqnarray}
For our example in Fig.~\ref{fig:Bpotential} where $Q=0.3\,$, the mass range for  a black hole to exist is $m \geq m_0 \approx 0.448$ providing  a lower bound  to the black hole radius $x \geq x_0 \approx 0.36\,$.

For very large distances ($x \gg |Q|$), the line element \eqref{BAdS2_potential} coincides with the conventional SAdS black hole where
\begin{equation}
f(x) \approx 1-\frac{2m}{x}+x^2 \,,
\end{equation}
while near the origin ($x \ll |Q|$) the metric potential can be approximated by
\begin{equation} \label{BAdS3_potential}
f(x) \approx 1-  \frac{\Lambda_{\mathrm{eff}}}{3} x^2 \,,
\end{equation}
where
\begin{equation}
\Lambda_{\mathrm{eff}} = 3 \left(\frac{2m}{Q^3} - 1 \right) \,.
\end{equation}
Depending on the value of $m\,$, the origin can be seen as:
\begin{itemize}
\item a repulsive deSitter core $(\Lambda_{\mathrm{eff}}>0)\,$ if $m > Q^3/2\,$;
\item an attractive AdS core $(\Lambda_{\mathrm{eff}}<0)\,$ if $m < Q^3/2\,$;
\item a local Minkowski core $(\Lambda_{\mathrm{eff}}=0)\,$ with no gravitational interaction if  $m = Q^3/2\,$.
\end{itemize}
In the case of an AdS core at the origin, the curve \eqref{BAdS3_potential} keeps growing and never reaches the x-axis, admitting no horizons. Therefore, black holes exist only when $m \geq m_0 \geq Q^3/2\,.$

\section{Chemitry of BAdS black hole}
\label{BAdS_chemistry}

\begin{figure}[h!] 
\begin{center}
\includegraphics[width=0.58 \textwidth]{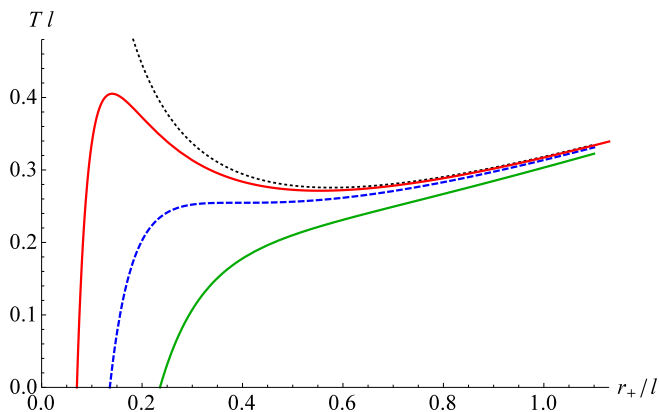}
\caption{ The temperature $\mathcal{T}=T l$ \textit{vs} the horizon $x_+=r_+/l$ for $Q=0.05$ (red solid curve) with a zero temperature remnant at $x_0=0.0702 \,$, for $Q=Q_c \approx 0.0987$ (blue dashed curve) with $x_0=0.136 $ and for $Q=0.18$ (green solid curve) with $x_0 = 0.236 $. The black dotted curve corresponds to the conventional SAdS solution ($Q=0$).}
\label{fig:Btemp}
\end{center}
\end{figure}

In order to see if there are any chemical aspects, we have to derive all possible thermodynamic variables that describe our system. As shown and discussed in \cite{NiT11,Nic10}, we can extend the known thermodynamic relations to the case in which a specific microscopic structure of the quantum spacetime is prescribed. Furthermore, the metric \eqref{BAdS_potential} appears an asymptotic behaviour at infinity, allowing us  to define a conserved mass-energy with respect to the event horizon without taking into account the inner horizon. The relation $f(r_+)=0\,$ gives  the black hole mass 
\begin{equation} \label{B_mass}
M=H= \frac{(l^2+r_+^2) (q^2 + r_+^2)^{3/2}}{2l^2r_+^2} \,,
\end{equation}
which also represents the chemical enthalpy $H$ of the system in the extended phase space. From \eqref{BAdS_potential}, we find that the temperature  reads
\begin{equation} \label{B_temp0}
T = \frac{3r_+^4 + l^2(r_+^2-2q^2)}{4\pi l^2 r_+ (q^2 + r_+^2)} \,
\end{equation}
and re-expressing it in dimensionless units through \eqref{dim_var}, we get
\begin{equation} \label{B_temp}
\mathcal{T} = \frac{3x_+^4+x_+^2-2Q^2}{4\pi x_+ (Q^2 + x_+^2)} \,.
\end{equation}
The temperature has to be non-negative, i.e., $3x_+^4+x_+^2-2Q^2 \geq 0\,$, leading to the condition $x_+ \geq \frac{\sqrt{-1+ \sqrt{1+24 Q^2}}}{\sqrt{6}} = x_0\,$. A plot of  the temperature \eqref{B_temp} is illustrated in Fig.~\ref{fig:Btemp}.
For values below the critical 'charge' ($Q<Q_c$), the temperature appears two extrema; one maximum and one minimum (red solid curve in Fig.~\ref{fig:Btemp}). For $Q=Q_c$ the two extrema merge at one inflexion point (blue dashed curve in Fig.~\ref{fig:Btemp}), while for $Q>Q_c$ the temperature is a monotonically increasing function of the radius (green solid curve in Fig.~\ref{fig:Btemp}).
The entropy of the black hole can be evaluated from thermodynamic arguments, $T = \left( \frac{\partial M}{\partial S} \right)_{P,q} $. Since no black hole exists below $r_0\,,$ we integrate the above relation from the minimum radius $r_0$ up to the horizon radius $r_+\,$ and  we retrieve the  entropy
\begin{equation} \label{B_s}
S= \int \limits^{r_+}_{r_0} \mathrm{d} r \ 2\pi r \left( 1 + \frac{q^2}{r^2} \right)^{3/2} = \pi r^2 \left[ \left( 1 - \frac{2q ^2}{r^2} \right) \sqrt{1+\frac{q^2}{r^2}} + \frac{3q ^2}{r^2} \ln \left( r + \sqrt{r^2 + q ^2} \right)   \right] \Bigg|_{r_0}^{r_+}   \,,
\end{equation}
which concludes to the usual area law ($S \approx \pi r_+^2$) up to a logarithmic corrections in the classical limit.

The cosmological term allows for the definition of the pressure \eqref{press} of the system along with its conjugate quantity,  the  black hole volume $V=\left( \frac{\partial M}{\partial P} \right)_{S,q}.$
Moreover, one may consider that the black hole mass is  a function  of three thermodynamic variables instead of two, i.e., $M=M(S,P,q)\,.$ This way, we include  thermodynamic contributions of the minimal length  $q$ inside the 1st law and the Smarr relation.  Here we take into account the above form of the entropy   by using \eqref{B_s} in the forthcoming calculations. 
The differential of the mass will give 
\begin{equation}
\mathrm{d} M = \left( \frac{\partial M}{\partial S} \right)_{P,q} \mathrm{d} S + \left( \frac{\partial M}{\partial P} \right)_{S,q} \mathrm{d} P + \left( \frac{\partial M}{\partial q} \right)_{S,P} \mathrm{d} q\,,
\end{equation}
 with the quantity $\varphi = \left( \frac{\partial M}{\partial q} \right)_{S,P}$ being the conjugate variable of $q\,$. Thus the new 1st law  for our system reads
\begin{equation} \label{new_1st_law}
\mathrm{d} M = T \mathrm{d} S + V \mathrm{d} P + \varphi \mathrm{d} q \,.
\end{equation}
 Below we gather all the desired thermodynamic quantities of the BAdS black hole that satisfy  \eqref{new_1st_law}:
\begin{eqnarray}
M &=&  \frac{(l^2+r_+^2) (q^2 + r_+^2)^{3/2}}{2l^2r_+^2} \,, \ \ \  \varphi = \frac{3q (l^2+r_+^2) \sqrt{q^2 + r^2_+}}{2 l^2 r_+^2} \,,
\\ T &=& \frac{3r_+^4 + l^2 (r_+^2-2q^2)}{4\pi r_+ l^2 (q^2 + r_+^2)} \,, \ \ \ V = \frac{4\pi r_+^3}{3} \left(  1 + \frac{q^2}{r_+^2}  \right)^{3/2} \,,  \  \ \ P = \frac{3}{8\pi l^2}\,, \label{Thd_var_2}\\   S &=& \pi r^2 \left[ \left( 1 - \frac{2q ^2}{r^2} \right) \sqrt{1+\frac{q^2}{r^2}} + \frac{3q ^2}{r^2} \ln \left( r + \sqrt{r^2 + q ^2} \right)   \right] \Bigg|_{r_0}^{r_+}  \,. \label{Thd_var_3}
\end{eqnarray}
In the classical limit ($r_+ \gg q$) where the length $q$ is considered negligible, all the above quantities coincide with the conventional variables of a SAdS black hole, retrieving all the well-known thermodynamic phenomenology of SAdS spacetime \cite{KuM17}.

For the new Smarr relation, we use Euler’s theorem for homogeneous functions \cite{KRT09}, such as the black hole mass in our case, which provides us with a route between the 1st law of black hole mechanics and the Smarr formula for stationary black holes. Since  $q$ has dimensions of length, the scaling argument  will give
\begin{equation} \label{new_smarr}
M=2T S' - 2P V+ \varphi q \,.
\end{equation}
The entropy $S'\,,$ resulting from \eqref{new_smarr}, reads
\begin{equation}
S' = \pi r_+^2 \left( 1 + \frac{q^2}{r_+^2} \right) ^{3/2}
\end{equation}
and is different from the entropy \eqref{B_s} that satisfies the 1st law, i.e., $S \neq S'\,$. Therefore, the 1st law is inconsistent with the Smarr formula for the BAdS black hole  at distances where $q$ is comparable with the event horizon (see also \cite{Bre05}), unless we are working in the large-distance regime ($r_+ \gg q$) where corrections of $q$ can be ignored and the two entropies coincide, satisfying that way the usual area law, i.e., $S' = S \approx \pi r_+^2\,$.

\begin{figure}[h!] 
\begin{center}
\includegraphics[width=0.58 \textwidth]{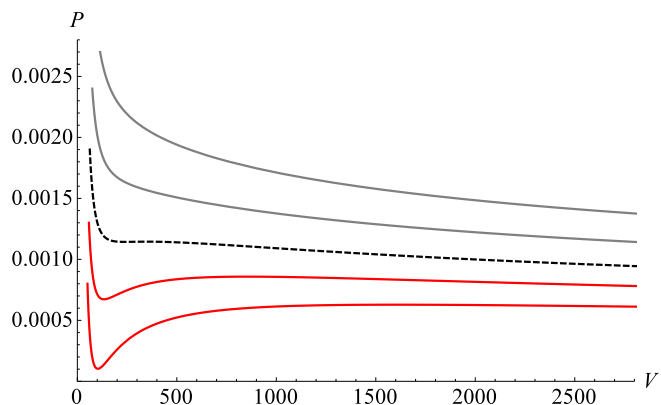}
\caption{ The isotherms of  BAdS black hole on the $P-V \,$ plane are displayed for $q=1\,$. The red solid curves stand for $T<T_c$, the black dashed curve for $T=T_c \approx 0.025$ and the gray solid curves for $T>T_c\,.$ }
\label{fig:Beos2}
\end{center}
\end{figure}

As for the physical thermodynamic interpretation of $q$ and $\varphi\,$, we can connect them with an extra pressure and a conjugate volume term respectively. Specifically, based on the discussion of \cite{MiX17,Lia17}, we could interpret the quantity 
\begin{equation}
P_q=-\frac{1}{8\pi q^2}
\end{equation}
as the pressure resulting from the existence of the minimal length $q\,$, in the same way the pressure $P$ results from the existence of $l\,,$ and the quantity 
\begin{equation}
V_q= \left( \frac{\partial M}{\partial P_q} \right) _{S,P} = \frac{6\pi q^5}{r_+^2} \left(1+ \frac{r_+^2}{l^2} \right) \sqrt{1+ \frac{r_+^2}{q^2}} 
\end{equation}
 as its conjugate variable having dimensions of volume. The  pressure $P_q$ should be negative, namely the tension of the self-gravitating droplet of anisotropic fluid doing work to the thermodynamic system by pushing against gravitational collapse. Therefore, the two pressures $P$ and $P_q$ have opposite effects on the system and, hence, opposite signs. Also, the volume $V_q$ turns out to be positive this way (for more details on the meaning of $P_q$ and $V_q$ see the aforementioned papers \cite{MiX17,Lia17}). 
The pair of conjugate variables ($P_q, V_q$) is used in the physical system  and is expected to appear in the new Smarr relation and the 1st law as an extra
pressure-volume term:
\begin{eqnarray}
M &=& 2 T S' - 2P V- 2 P_q V_q \,, \\
\mathrm{d} M &=& T \mathrm{d} S + V \mathrm{d} P + V_q \mathrm{d} P_q \,.
\end{eqnarray}
Combining now the expressions of the temperature $T$ and the volume $V$ with the pressure $P$ given by \eqref{Thd_var_2}, we find that the equation of state  $P=P(V,T)$ for the  BAdS black hole is given by
\begin{equation} \label{Beos2}
P=\frac{12q^2 + \left( \frac{6V}{\pi} \right)^{2/3} \left( -1+ 2\pi T \sqrt{\left( \frac{6V}{\pi} \right)^{2/3} - 4q^2 } \right)   }{2\pi \left( \left( \frac{6V}{\pi} \right)^{2/3} - 4q^2 \right)^2  }  \,.
\end{equation}
As can be seen in Fig.~\ref{fig:Beos2}, the isotherms of \eqref{Beos2} appear a characteristic unstable branch similar to Van der Waals theory when $T<T_c\,$,  with all thermodynamic quantities ($P,V,T$) of the BAdS black hole  identified with the same quantities of the Van der Waals gas.
The unstable branch resulting from \eqref{Beos2} can  be investigated through the two specific  heats and the Gibbs free energy $G(T,P)\,.$ The specific heats indicate the local stability of the system and divergences of them  signal the existence of a phase transition, while the Gibbs energy indicates the global stability of the system and the global minimum of $G$ corresponds to the preferred state. Having clear interpretation of the black hole volume and the pressure, we can  define the two specific heats; one with constant volume $C_V=\frac{1}{T} \left( \frac{\partial S}{\partial T} \right)_{V,q}=0 \,,$ which turns out to be zero in our case and one with constant pressure  $C_P = \frac{1}{T} \left( \frac{\partial S}{\partial T} \right)_{P,q}$ which reads
\begin{equation} \label{BAdSCp}
C_P = \frac{2\pi  (3r_+^4 -2q^2 l^2+l^2 r_+^2)(q^2+r_+^2)^{5/2}}{3r_+^5 (3q^2+r_+^2)+l^2 r_+ (2q^4+7q^2 r_+^2 -r_+^4)} \,.
\end{equation}
The Gibbs energy can be evaluated through the relation $G=M-T S\,$, whose expression is omitted here.
\begin{figure}[!h] %
\centering
\subfigure[The specific heat $C_P$ \textit{vs} the horizon $r_+\,$. ]{%
\label{fig:B4Cp}%
\includegraphics[height=1.58in]{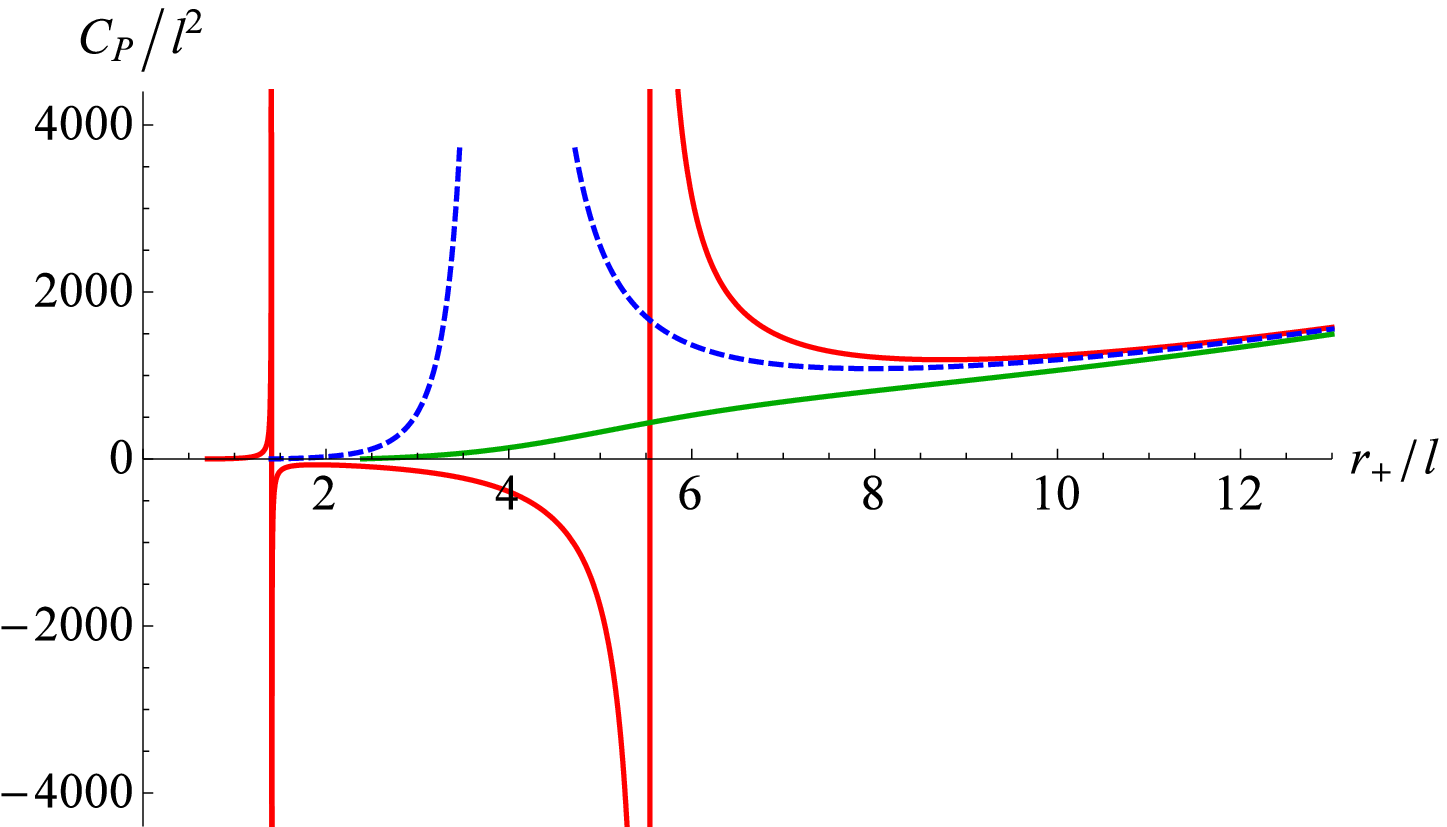}}%
\qquad
\subfigure[The Gibbs energy $G$ \textit{vs} the horizon $r_+\,$.]{%
\label{fig:B4G}%
\includegraphics[height=1.58in]{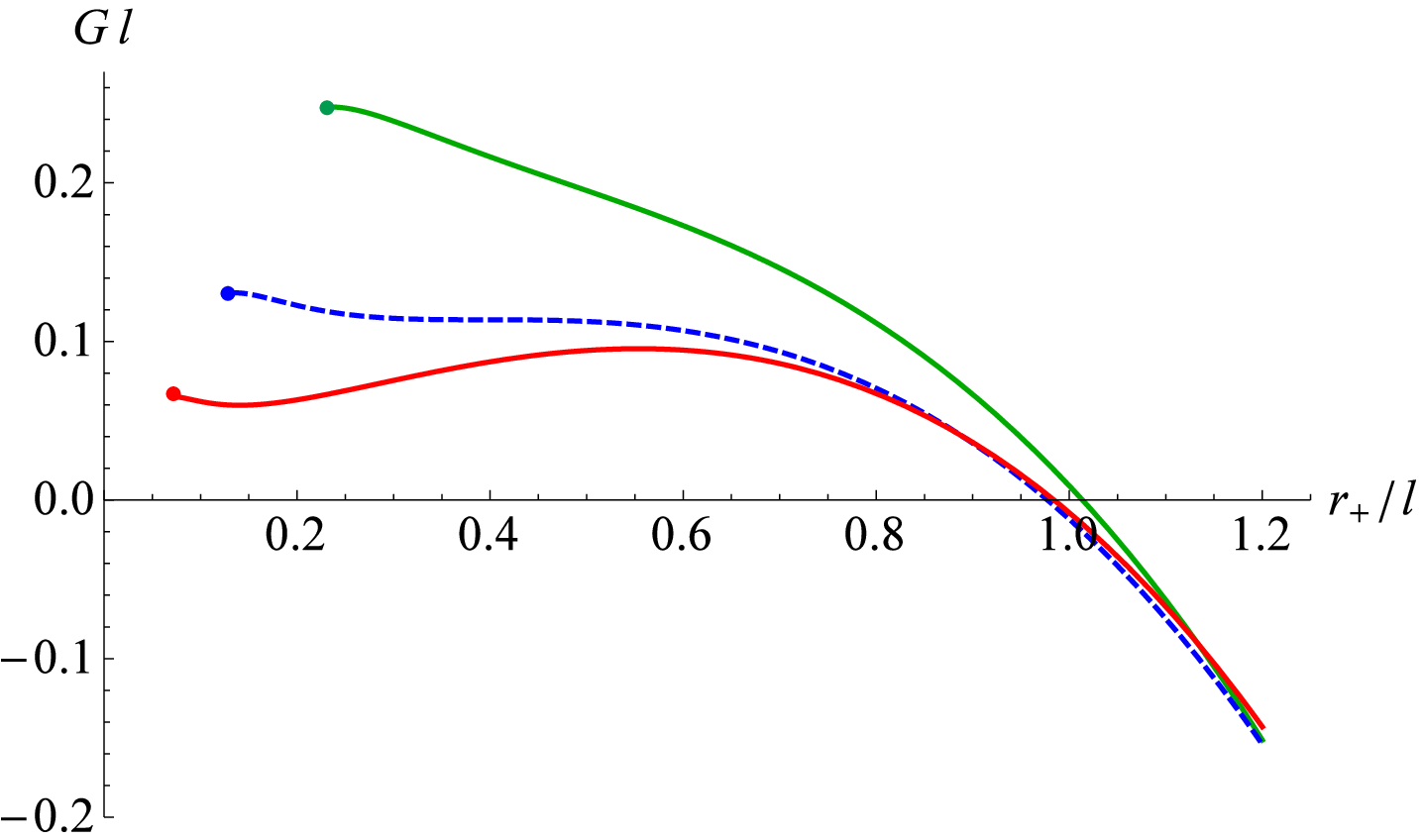}}%
\caption{The plots are displayed in $l$ units for $Q=0.05$ (red solid curves), for $Q=Q_c \approx 0.0987$ (blue dashed curves) and for $Q=0.18$ (green solid curves).}
\label{fig:B4CpG}
\end{figure}
We plot $C_P$ and $G$  in Fig.~\ref{fig:B4CpG}. As can be seen, we can distinguish 3 different cases:
\begin{itemize}

\item For $Q=0.05<Q_c  $ we get three branches (red solid curves in Fig.~\ref{fig:B4CpG}); one small locally stable BAdS black hole $(C_P>0)$ for $0.0702 \lesssim r_+ / l \lesssim 0.14 \,,$ one intermediate unstable black hole ($C_P<0$) for $ 0.14 \lesssim r_+ / l \lesssim 0.55 \,$ and one large stable black hole $(C_P>0)$ for  $r_+ / l \gtrsim 0.55\,.$ The divergences of $C_P$ occur at the two phase transition points $r_s/l \simeq 0.14$  and $r_l/l \simeq 0.55$ where the temperature appears a maximum and a minimum respectively (Fig. \ref{fig:Btemp}) while the Gibbs energy appears a minimum and a maximum respectively (Fig.~\ref{fig:B4G}). Therefore, a first order phase transition seems to take place between small/large stable black hole.

\item For $Q=Q_c \approx 0.0987\,$ we get two branches (blue dashed curves in Fig.~\ref{fig:B4CpG}); one small and one large stable black hole  coexisting at one inflexion point  $r_i /l \approx0.392 \,$ where $C_P \rightarrow \infty\,.$ 

\item For $Q=0.18>Q_c$ no phase transition occurs and so there exists one thermodynamically stable black hole (green solid curves in Fig.~\ref{fig:B4CpG}). The preferred states are those with larger horizons because the Gibbs energy decreases with the increase of $r_+\,$.
\end{itemize}
The $P-V$ criticality of BAdS black hole  can be  checked from the condition
\begin{equation} \label{crtical_eq}
\frac{\partial P}{\partial V} = 0 = \frac{\partial ^2 P}{\partial V^2} \,.
\end{equation}
The solution of \eqref{crtical_eq} yields a critical point with 
\begin{equation}
P_c \simeq \frac{0.0012}{q^2}\,, \ \ \ \ V_c \simeq 287.44 q^3\,, \ \ \ \ T_c \simeq \frac{0.0251}{q } \,,
\end{equation}
 at which one cannot distinguish between small/large black hole. The universal constant
 \begin{equation}
 \varepsilon = \frac{P_c V_c^{1/3}}{T_c} \approx 0.31
 \end{equation}
characterizes the thermodynamic system and  for the BAdS black hole is slightly different from the value $0.375$ of the Van der Waals gas, making the analogy between them even stronger but not identical.

Last but not least, we evaluate the critical exponents $\alpha,\beta,\gamma,\delta,\,$ which describe the behaviour of physical quantities near the critical point \cite{GMK13}. Defining the dimensionless quantities 
\begin{equation} \label{dimensionless_crit}
p=\frac{P}{P_c} \,, \ \ \ \ v=\frac{V}{V_c} \,  \ \ \ \ \tau=\frac{T}{T_c} \,,
\end{equation}
we extract the \textit{law of corresponding states}   by substituting the above expressions \eqref{dimensionless_crit} into the equation of state \eqref{Beos2}:
\begin{equation} \label{Blocs}
p=\frac{0.354 + v^{2/3} \left(  -1.978 + 0.312 \ \tau \sqrt{-4+67.045 \ v^{2/3}} \right) }{\left( 0.06 - v^{2/3}\right) ^2} \,.
\end{equation}
The above law is universal and valid under more general assumptions than eq.~\eqref{Beos2}. Expanding around the critical point by introducing the quantities
\begin{equation} \label{dimensionless_crit2}
t=\tau - 1,\, \ \ \ \ \omega = v - 1 \,,
\end{equation}
we can approximate \eqref{Blocs} as 
\begin{equation} \label{Blocs_approx}
p \approx 0.965 + 2.802 t -1.112 \omega t - 0.054 \omega^3 \,.
\end{equation}
The critical exponents can be found through the relations:
\begin{eqnarray} \label{crit_a}
C_V &=& T \frac{\partial S}{\partial T} \Big |_V \propto |t|^{-\alpha} \,, \\ \label{crit_b}
\eta &=& V_l - V_s \propto |t|^{\beta} \,, \\ \label{crit_c}
\kappa_T &=& - \frac{1}{V} \frac{\partial V}{\partial P} \Big |_T \propto |t|^{-\gamma} \,, \\ \label{crit_d}
|P-P_c|_{T=T_c}  & \propto &  |V-V_c|^{\delta} \,,
\end{eqnarray}
where $V_s$ and $V_l$ are the volumes of the small and large black hole respectively and $\kappa_T$ is the compressibility coefficient. 

The first exponent $\alpha$ describes the behaviour of the specific heat with constant volume and in our case is zero, $\alpha=0\,,$ because $C_V=0$ and so there is no dependence on $|t|\,.$ 

The second exponent $\beta$ can be evaluated by using \eqref{Blocs_approx}, as well as the known \textit{Maxwell's area law} 
\begin{equation} \label{Mal}
\oint  V \mathrm{d} P =0 \,,
\end{equation}
which replaces the oscillating part of the isotherm with an isobar in a canonical ensemble, implying that the pressure of the system remains constant during the transition instead of oscillating. The differential of \eqref{Blocs_approx} for fixed $t$ gives
\begin{equation} \label{dp}
\mathrm{d} p = -(0.162 \omega ^2 +1.112 t) \mathrm{d} \omega \,.
\end{equation}
From \eqref{Blocs_approx}, \eqref{Mal} and \eqref{dp} we get the following system:
\begin{eqnarray}
0 &=& \int \limits ^{\omega _l}_{\omega _s} \ \mathrm{d} \omega  (0.162 \omega ^2 +1.112 t) \omega  \,, \\ \nonumber
p &=& 0.965 + 2.802 t -1.112 \omega _s t - 0.054 \omega _s^3  \\
&=& 0.965 + 2.802 t -1.112 \omega_l t - 0.054 \omega _l^3  \,.
\end{eqnarray}
The solutions of the above two equations are
\begin{equation}
\omega_l=-\omega_s \,, \ \ \ \ \omega_l= - 0.5 \omega_s + 0.097 \sqrt{-81 \omega ^2_s-2224 t} \,,
\end{equation}
giving
\begin{equation}
\omega_l=-\omega_s \simeq 4.547 \sqrt{-t} \,.
\end{equation}
Thus eq.~\eqref{crit_b} can be written as
\begin{equation}
\eta=V_l-V_s = V_c (\omega_l-\omega_s) \simeq 9.1 V_c \sqrt{-t} \,,
\end{equation}
identifying the exponent $\beta$ with the value $\beta=1/2\,.$

For the third exponent $\gamma\,,$ we substitute  \eqref{dimensionless_crit}, \eqref{dimensionless_crit2} and \eqref{dp} into the  compressibility \eqref{crit_c}, retrieving
\begin{equation}
\kappa_T = - \frac{1}{(\omega+1)P_c} \frac{\mathrm{d} \omega }{ \mathrm{d} p} \propto \frac{1}{6.86t } \,.
\end{equation}
Therefore, the exponent $\gamma$ equals to unity, $\gamma=1\,.$

For the fourth exponent $\delta\,$, we set $T=T_c$ having that way $t=0\,$ and so eq.~\eqref{crit_d} takes the form
\begin{equation}
p-1=0.965-0.054 \omega^3 \,,
\end{equation}
identifying the exponent $\delta$ with the value $\delta=3\,.$ One sees that the critical exponents of  BAdS black hole $(\alpha,\beta,\gamma,\delta)=(0,1/2,1,3)$ are identical to those of a Van der Waals gas.

\section{Conclusions}
\label{concl}

We have presented a geometric and thermodynamic analysis of the Bardeen black hole inside an extended AdS phase space. From a geometrical point of view, we derived the line element of  BAdS black hole where the center is described by a regular core and we showed that such a black hole can have two, one or no horizons depending on the value of the black hole mass relative to the mass of the minimum black hole set by the minimal cut-off length $q$ and the AdS radius $l\,$. From a thermodynamical point of view, we considered that the cosmological constant plays the role of the positive pressure of the system and, after having calculated the desired thermodynamic variables, we derived the black hole equation of state which appears an unstable branch similar to that of a Van der Waals gas. The resultant forms of the specific heat and the Gibbs free energy indicate a first order phase transition between small/large black hole reminiscent of the liquid/gas coexistence. Moreover, we considered the minimal length $q$ as a thermodynamic variable and we saw that the new Smarr relation is inconsistent with the 1st law at distances where the event horizon is comparable with $q\,$. Finally, we showed that the system appears criticality and we derived the critical exponents of the black hole which are identical to  those of a Van der Waals gas, making the analogy between them even stronger.

\subsection*{Acknowledgments}

The author would like to thank Piero Nicolini for suggesting the topic of black hole chemistry and for providing valuable comments to the
manuscript.

\end{document}